\documentclass[article,amsmath,11pt,amssymb,floatfix,superscriptaddress,
showpagenumber,showpacs,nofootinbib]{revtex4}
\usepackage{graphicx}
\usepackage{epsfig}
\usepackage{bm}
\usepackage{amsfonts}

\input epsf

\begin{document}

\title{\Large Viscous extended holographic Ricci dark energy in the framework of standard Eckart theory}

\author{Surajit Chattopadhyay}
\email{surajitchatto@outlook.com, surajcha@associates.iucaa.in}
\affiliation{ Pailan College of Management and Technology (MCA Division), Bengal
Pailan Park, Kolkata-700 104, India.}

\date{\today}

\begin{abstract}
In the present work we report a study on the viscous extended holographic Ricci dark energy (EHRDE) model under the assumption of existence of bulk viscosity in the linear barotropic fluid and the EHRDE in the framework of standard Eckart theory of relativistic irreversible thermodynamics and it has been observed that the non-equilibrium bulk viscous pressure is significantly smaller than the local equilibrium pressure. We have studied the equation of state (EoS) parameter and observed that the EoS behaves like ``quintom" and is consistent with the constraints set by observational data sets from SNLS3, BAO and Planck + WMAP9 + WiggleZ measurements in the reference S. Kumar and L. Xu, \emph{Phys. Lett. B},\textbf{737}, 244 (2014). Analysis of statefinder parameters has shown the possibility of attainment of $\Lambda$CDM phase under current model and at the same time it has been pointed out that the for $z=0$ i.e. current universe, the statefinder pair is different from that of $\Lambda$CDM and the $\Lambda$CDM can be attained in a later stage of the universe. An analysis of stability has shown that although the viscous EHRDE along with viscous barotropic is classically unstable in the present epoch, it can lead to a stable universe in very late stage. Considering an universe enveloped by event horizon we have observed validity of generalized second law of thermodynamics.
\end{abstract}

\maketitle

\maketitle
\section{Introduction}

Independent studies by Riess et al. \cite{S27} and Perlmutter et al. \cite{S271} of high-redshift supernavoe search team and supernovae cosmology project team respectively reported that the current universe is expanding with acceleration. Subsequent observational studies including large-scale structure (LSS) and the cosmic microwave background (CMB) have further confirmed the accelerated expansion (see \cite{Luca}). In order to have this accelerated expansion there must be something to overcome the effect of gravity. ``Dark energy" (DE), an exotic matter characterized by negative pressure and having equation of state (EoS) parameter $w=p/\rho<-1/3$, is believed to be responsible for this accelerated expansion. However, its exact nature is yet to be known. Different candidates for DE have been proposed till date with varying behaviour of EoS parameter. The simplest candidate is the cosmological constant $\Lambda$ with $w_{\Lambda}=-1$ (reviewed in \cite{carroll,sahni1,sahni2,sahni3}). Although it is consistent with observations \cite{carroll, lambda1, bits}, Alam et al. \cite{sahni4} argued that the current accelerating epoch being unlikely to be unique one may suggest a time dependent form of DE. The time dependent candidates include tachyon field \cite{DE1,DE11,DE111}, quintessence \cite{quint1,quint2,cai}, phantom \cite{phantom1,phantom2,phantom3,phantom4}, hessence \cite{hessence1,hessence2,hessence3}, k-essence \cite{kess1,kess2,kess3}, Chaplygin gas \cite{DE2,DE3,DE4,DE5}, holographic dark energy \cite{3,holo1,holo2,holo3} etc. Some remarkable reviews on DE include \cite{copeland-2006,frieman,bambareview,DErev,nojirireview}. It is indicated by the most recent CMB observations that DE accounts for about three fourths of the total mass energy of the Universe \cite{threefourth}. Different reconstruction schemes for DE have also been proposed till date e.g. \cite{reconst1,reconst2,reconst3}.

Inspired by ``holographic principle" Li\cite{3} proposed ``Holographic dark energy (HDE)" having density $\rho_{\Lambda}=3c^2M_p^2L^{-2}$, where $L$ is the infrared cut-off. Works on HDE include \cite{holo1,holo2,holo3,holo4,holo5,holo6,holo7}. Different variants of HDE have been proposed in the literature. In the present work we consider a special form of HDE \cite{EHRDE1} dubbed as ``extended holographic Ricci dark energy" (EHRDE) \cite{EHRDE1}. Its density has the form
\begin{equation}\label{RD}
\rho_{DE}=3 M_p^2\left(\alpha H^2+\beta \dot{H}\right)
\end{equation}
where the upper dot represents derivative with respect to cosmic time $t$, $M_p^2$ is the reduced Planck mass, $\alpha$ and $\beta$
are constants to be determined. Wang and Xu
\cite{2} found the best-fit values in order to make this cutoff to
be consistent with observational data as
$\alpha=0.8502^{+0.0984+0.1299}_{-0.0875-0.1064}$ and
$\beta=0.4817^{+0.0842+0.1176}_{-0.0773-0.0955}$. In the current work we shall take
$\alpha= 0.98$ and $\beta=0.37$.

Nojiri and Odintsov \cite{nojiri5,nojiri6} developed cosmological models treating dark energy and dark matter as imperfect fluids with unusual equation of state, where viscous fluids are just one particular case. Important role of bulk and shear viscosity in the early phase of the evolution of the universe has been emphasized in \cite{vcg}. Chimento et al \cite{vis1} reported that a  combination of cosmic fluid with bulk dissipative pressure and quintessence matter can derive accelerated expansion of the universe that involves a sequence of important dissipative processes that including GUT phase transition at $t\approx 10^{-34} s$ and a temperature of about $T\approx 10^{27}K$ \cite{vcg}. First attempts towards creating a theory of relativistic dissipative fluids were made by Eckart \cite{eckart} and Landau and Lifshitz \cite{landu}. Israel and Stewart \cite{israel} developed a relativistic second-order theory. Nojiri and Odintsov \cite{nojiriadd2} made a time dependent viscosity consideration to DE by considering EoS with inhomogeneous, Hubble parameter dependent term. Brevik et al. \cite{nojiriadd3} discussed entropy of DE in the framework of holographic Cardy-Verlinde formula and in a relatively recent work, Brevik et al. \cite{ nojiriadd4} derived a formula for the entropy for a multicomponent coupled fluid that could relate the entropy of a closed FRW universe to the energy contained in it together with its Casimir energy.  Brevik et al. \cite{apss} investigated interacting dark energy and dark matter in flat FRW universe with examples of Little Rip, Pseudo Rip, and bounce cosmology and expressed bulk viscosity as function of Hubble parameter and time.

Plan of the present paper is as follows: In Section II we have presented the cosmological consequences of existence of bulk viscosity in the linear barotropic fluid and the EHRDE in the framework of standard Eckart theory of relativistic irreversible thermodynamics through the study of reconstructed Hubble parameter, EoS parameter and statefinder diagnostics. In Section III we have presented the stability analysis and in Section IV we have examined the validity of generalized second law of thermodynamics under the assumption that the universe is enveloped by event horizon. We have concluded in Section V. 

\section{Eckart approach}

For an homogeneous and isotropic flat universe the FRW metric is 
\begin{equation}
ds^2=-dt^2+a(t)^2\left(dr^2+r^2(d\theta^2+\sin^2\theta d\phi^2)\right)
\end{equation}
where, $a(t)$ is the scale factor, $t$ is the cosmic time. According to the first order thermodynamic theory of Eckart \cite{eckart} the field
equations in the presence of bulk viscous stresses are 
\begin{equation}
\left(\frac{\dot{a}}{a}\right)^2=H^2=\frac{\rho}{3}
\end{equation}
\begin{equation}
\frac{\ddot{a}}{a}=\dot{H}+H^2=-\frac{1}{6}\left(\rho+3P_{eff}\right)
\end{equation}
where, $P_{eff}=p+\Pi$ in which $\Pi$ is the bulk viscous pressure and
\begin{equation}
\Pi=-3H\xi
\end{equation}
where, $\xi$ is the bulk viscosity coefficient. The condition $\xi>0$ ensures production of positive entropy. Cataldo et al. \cite{cataldo} applied Eckart approach to study the effect of bulk viscosity on big rip singularity in the flat FRW cosmologies. Feng and Li \cite{feng} studied the viscous Ricci dark energy model under the assumption that there is bulk viscosity in the Ricci dark energy and linear barotropic fluid. Here, we extend the study of \cite{feng} to extended holographic Ricci dark energy (EHRDE). At this juncture, it may be noted that Eckart approach suffers from some limitations. In this theory, dissipative perturbations propagate at infinite speeds and the equilibrium states in the theory are unstable \cite{titus}. Israel and Stewart \cite{israel} theory, a causal and stable one, can obtain the Eckart theory when the relaxation time goes to $0$. In spite of the said limitations, because of simplicity of approach, Eckart theory has been adopted widely in the study of the accelerating universe filled with bulk viscous fluid (e.g.\cite{cataldo,feng,titus}). In the present work, we consider that in addition to the viscous EHRDE the universe also contains barotropic fluid, and consequently the Friedman equation and the corresponding equations of motion take the form \cite{feng}
 \begin{equation}\label{21}
3H^2=\rho_{DE}+\rho_\nu
\end{equation}
and
\begin{eqnarray}
\rho_{DE}'=-3 (\rho_{DE}+p_{DE}+\Pi_{DE})\label{DE}\\
\rho_{\nu}'=-3(\nu \rho_{\nu}+\Pi_{\nu})
\end{eqnarray}
where, the barotropic fluid has the equation of state $p_{\nu}=(\nu-1)\rho_{\nu};~(0\leq \nu \leq2)$ and $\Pi_{DE}$ and $\Pi_{\nu}$ are to be defined later based on the physically natural fact that the bulk viscosity is proportional to the fluid's velocity vector. The prime denotes the derivative with respect to $x=\ln a$. From Eq.(\ref{21}) we can write
\begin{equation}
3H^2=3 \left(\alpha H^2+\frac{\beta}{2}\frac{dH^2}{dx}\right)+\rho_{\nu}
\end{equation}
that implies
\begin{equation}
\rho_{\nu}=3 \left((1-\alpha)H^2- \frac{\beta}{2}\frac{dH^2}{dx}\right)
\end{equation}
and from barotropic equation of state we get
\begin{equation}
p_{\nu}=3(\nu-1) \left((1-\alpha)H^2- \frac{\beta}{2}\frac{dH^2}{dx}\right)
\end{equation}
Subsequently, we can have
\begin{equation}\label{pinu}
\Pi_{\nu}=-3\sqrt{3}\tau_{\nu} y= 3 (1-\alpha ) \nu  y+(1- \alpha -\frac{3}{2} \beta  \nu ) y'-\frac{\beta}{2}  y''
\end{equation}
where $y=H^2$ and the primes indicate derivative with respect to $x=\ln a$. We can rewrite Eq.(\ref{pinu}) as
\begin{equation}\label{pinu1}
\beta  y''-(2 (1- \alpha) -3 \beta  \nu ) y'-6( (1-\alpha ) \nu-\sqrt{3}\tau_{\nu})y=0
\end{equation}
solving which we get
\begin{equation}\label{ysquare}
H^2= C_1e^{x \Delta^{-}} +(H_0^2-C_1)e^{x\Delta^{+}}
\end{equation}
where
\begin{equation}\label{ddelta}
\Delta^{\mp}=\frac{1}{2\beta}\left(2-2 \alpha -3 \beta  \nu \mp\sqrt{(-2+2 \alpha +3 \beta  \nu )^2-4 \beta  \left(-6 \nu +6 \alpha  \nu +6\sqrt{3}\tau_{\nu} \right)}\right)
\end{equation}
and $H_0^2=C_1+(H_0^2-C_1)$. From Eq.(\ref{ddelta}) we understand that for real $\Delta^{\mp}$ we require
\begin{equation}
\sqrt{3}\tau_{\nu}\leq 6 (1-\alpha) \nu+\frac{(2(\alpha-1)+3 \beta\nu)^2}{4\beta}
\end{equation}
Thereafter, energy density of the viscous extended holographic RDE is
\begin{equation}\label{rhode}
\rho _{\text{DE}}=3 \left[\alpha  \left(C_1 e^{x \Delta ^-}+e^{x \Delta ^+} \left(-C_1+H_0^2\right)\right)+\frac{1}{2}
\beta  \left(C_1 e^{x \Delta ^-} \Delta ^-+e^{x \Delta ^+} \left(-C_1+H_0^2\right) \Delta ^+\right)\right]
\end{equation}
and that of the barotropic fluid is
\begin{equation}\label{rhonu}
\rho _{\nu }=-\frac{3}{2} \left[e^{x \Delta ^+} H_0^2 \left(-2+2 \alpha +\beta  \Delta ^+\right)+C_1 \left(2 \left(e^{x \Delta ^-}-e^{x \Delta ^+}\right)
(-1+\alpha )+e^{x \Delta ^-} \beta  \Delta ^--e^{x \Delta ^+} \beta  \Delta ^+\right)\right]
\end{equation}
Taking $\Pi_{DE}=-3\sqrt{3} \tau_{DE}H^2$ based on Eckart approach \cite{eckart} we have from Eq.(\ref{DE})
\begin{equation}
\begin{array}{c}
p_{DE}=-\frac{1}{2} C_1 \left(6 e^{x \Delta ^-} \alpha -6 e^{x \Delta ^+} \alpha -6 \sqrt{3} \left(e^{x \Delta ^-}-e^{x \Delta ^+}\right)
\tau _{\text{DE}}+e^{x \Delta ^-} (2 \alpha +3 \beta ) \Delta ^-\right.\\
\left.+e^{x \Delta ^-} \beta  \left(\Delta ^-\right)^2-2 e^{x \Delta ^+} \alpha  \Delta
^+-3 e^{x \Delta ^+} \beta  \Delta ^+-e^{x \Delta ^+} \beta  \left(\Delta ^+\right)^2\right)-
\frac{1}{2} e^{x \Delta ^+} \\
H_0^2 \left(-6 \sqrt{3}
\tau _{\text{DE}}+\left(3+\Delta ^+\right) \left(2 \alpha +\beta  \Delta ^+\right)\right)
\end{array}\label{pde}
\end{equation}
and hence the EoS parameter $w_{DE}$ is
\begin{equation}\label{eosD}
w_{DE}=\frac{p_{DE}}{\rho_{DE}}=-1+\frac{C_1 e^{x \Delta ^-} \left(6 \sqrt{3} \tau _{\text{DE}}-\Delta ^- \left(2 \alpha +\beta  \Delta ^-\right)\right)-e^{x \Delta
^+} \left(C_1-H_0^2\right) \left(6 \sqrt{3} \tau _{\text{DE}}-\Delta ^+ \left(2 \alpha +\beta  \Delta ^+\right)\right)}{3 C_1 e^{x \Delta
^-} \left(2 \alpha +\beta  \Delta ^-\right)-3 e^{x \Delta ^+} \left(C_1-H_0^2\right) \left(2 \alpha +\beta  \Delta ^+\right)}
\end{equation}
and
\begin{equation}
\begin{array}{c}\label{eos}
w_{total}=\frac{p_{DE}+p_{\nu}}{\rho_{DE}+\rho_{\nu}}=-1+\frac{1}{6} \left(6 \nu  (1-\alpha )+6 \sqrt{3} \tau _{\text{DE}}-\Delta ^+ \left(2 \alpha +3 \beta  \nu +\beta  \Delta ^+\right)-\right.\\
\left.\frac{C_1
e^{x \Delta ^-} \left(\Delta ^--\Delta ^+\right) \left(2 \alpha +3 \beta  \nu +
\beta  \left(\Delta ^-+\Delta ^+\right)\right)}{C_1 e^{x \Delta
^-}+e^{x \Delta ^+} \left(-C_1+H_0^2\right)}\right)
\end{array}
\end{equation}
Defining effective pressure $\tilde{p}=p_{DE}+p_{\nu}+\Pi_{DE}+\Pi_{\nu}$ we have
\begin{equation}
\begin{array}{c}\label{ptilda}
\tilde{p}=-e^{x \Delta ^+} H_0^2 \left(3+\Delta ^+\right)+C_1 \left(-3 e^{x \Delta ^-}+3 e^{x \Delta ^+}-e^{x \Delta ^-} \Delta ^-+e^{x
\Delta ^+} \Delta ^+\right)
\end{array}
\end{equation}

\begin{figure}[h]
\includegraphics[width=20pc]{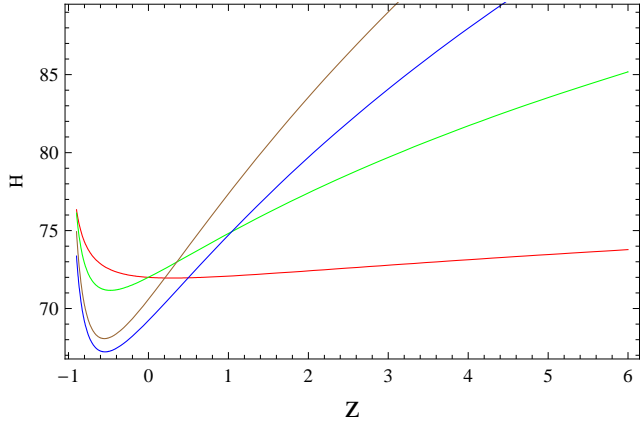}
\caption{\label{figH} Reconstructed Hubble parameter $H$ based on Eq.(\ref{ysquare}). }
\end{figure}

\begin{figure}
\centering
\begin{minipage}{.5\textwidth}
  \centering
  \includegraphics[width=18pc]{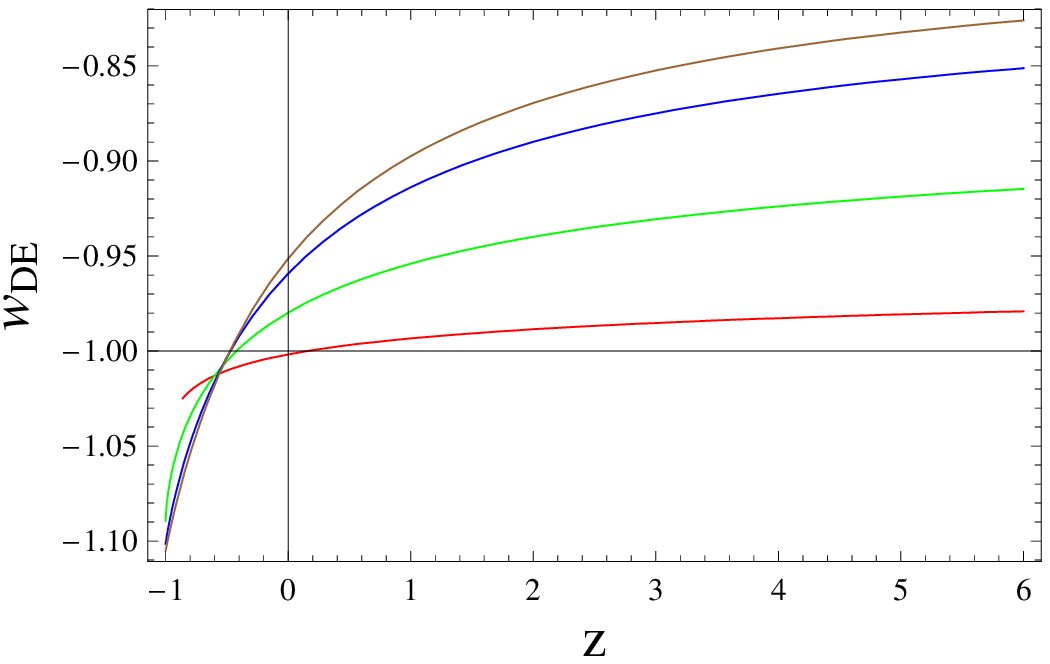}
  \caption{Plot of EoS parameter $w_{DE}$\\ based on Eq.(\ref{eosD}).}
  \label{figeosDE}
\end{minipage}%
\begin{minipage}{.5\textwidth}
  \centering
  \includegraphics[width=18pc]{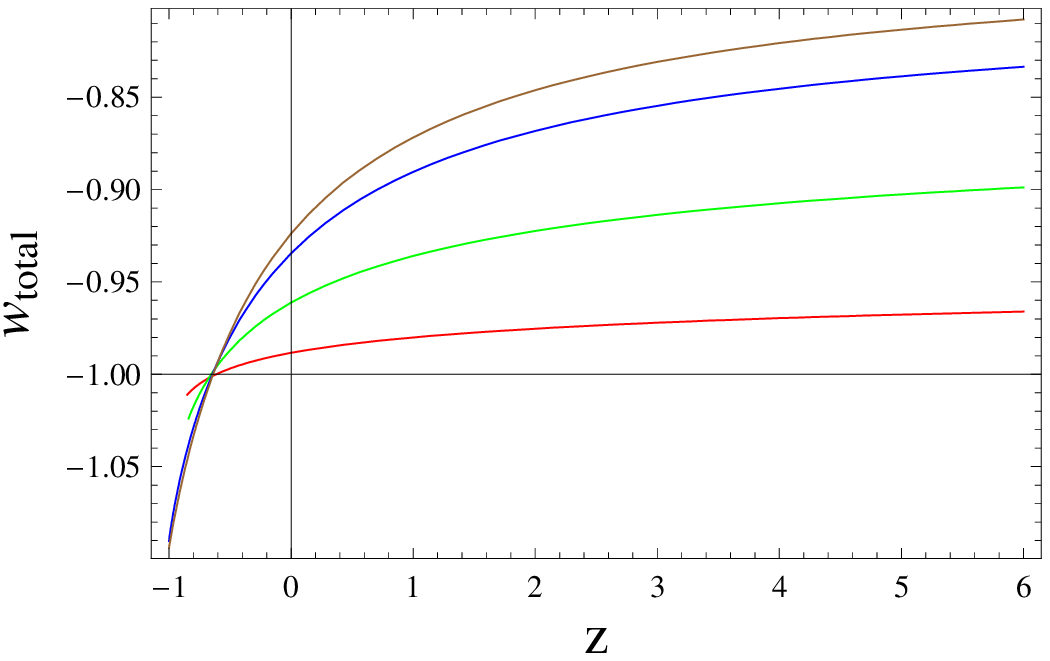}
  \caption{Plot of EoS parameter $w_{total}$\\ based on Eq.(\ref{eos}). }
  \label{figeostotal}
\end{minipage}
\end{figure}

In Fig.\ref{figH} we have plotted the reconstructed $H$ as a function of redshift $z$ based on Eq. (\ref{ysquare}). In this and subsequent plots black, red, green and blue lines will correspond to $\nu=0.15,~0.20,~0.25$ and $0.27$ respectively. The reconstructed Hubble parameter is found to be decreasing with evolution of the universe and this behaviour is consistent with the accelerated expansion of the current universe. However, in a later stage $-0.7\lesssim z\lesssim -0.25$, the reconstructed $H$ is found to start increasing. Thus, in a later stage $\dot{a}(t)$ may dominate $a(t)$.
Equation of state (EoS) parameter for the viscous EHRDE based on Eq.(\ref{eosD}) is presented in Fig. \ref{figeosDE} and we observe that a transition from $EoS>-1$ (quintessence) to $EoS<-1$ (phantom) is occurring at $z\approx 0.01$ i.e. in an earlier stage of the universe and as seen in Table \ref{tabeos} the current value of the EoS parameter for viscous EHRDE is favouring the $\Lambda$CDM model and the $w_{DE0}$ for the current model is consistent with results obtained by \cite{bits} through observational data sets from SNLS3, BAO and Planck+WMAP9+WiggleZ measurements. If we give a minute look at Fig.\ref{figeosDE} we can observe that for $\nu=0.15$ the crossing of phantom boundary is occurring at earlier stage i.e. $z\gtrsim 0$ and the current universe is in phantom era. However, for $\nu=0.25$ and $0.27$ the crossing of phantom divide $(EoS=-1)$ is occurring at later stage i.e. $z<0$ and the current universe is in quintessence era. Thus, for higher values of $\nu$ the transition from quintessence to phantom is getting delayed. However, irrespective of the values of $\nu$ the viscous EHRDE is behaving like ``quintom" i.e. transiting from quintessece to phantom. If we look at Fig.\ref{figeostotal} we can understand that the behaviour of $w_{total}$ is largely similar to that of $w_{DE}$ as far as the ``quintom" behaviour is concerned. However, the transition to phantom is occurring in a later stage $z\approx -0.6$ and this is true for all values of $\nu$. Hence, for $w_{total}$ the time point at which the universe is transiting to phantom from quintessence is not influenced by the value of $\nu$.

\begin{table}[h!]
  \centering
  \caption{$w_{total}$ and $w_{DE}$ at $z=0$ for different values of $\nu$.}
  \label{tabeos}
  \begin{tabular}{c|c|c|c|c}
    \hline\hline
    EoS parameter & $\nu=0.15$ & $\nu=0.20$ & $\nu=0.25$ & $\nu=0.27$\\
    \hline
   $w_{total0}$ &-0.988403& -0.961182& -0.934464& -0.923836\\
    \hline
   $w_{DE0}$ & -1.0018& -0.979765& -0.95919& -0.9513\\
   \hline
  \end{tabular}
\end{table}

\begin{figure}
\centering
\begin{minipage}{.5\textwidth}
  \centering
  \includegraphics[width=18pc]{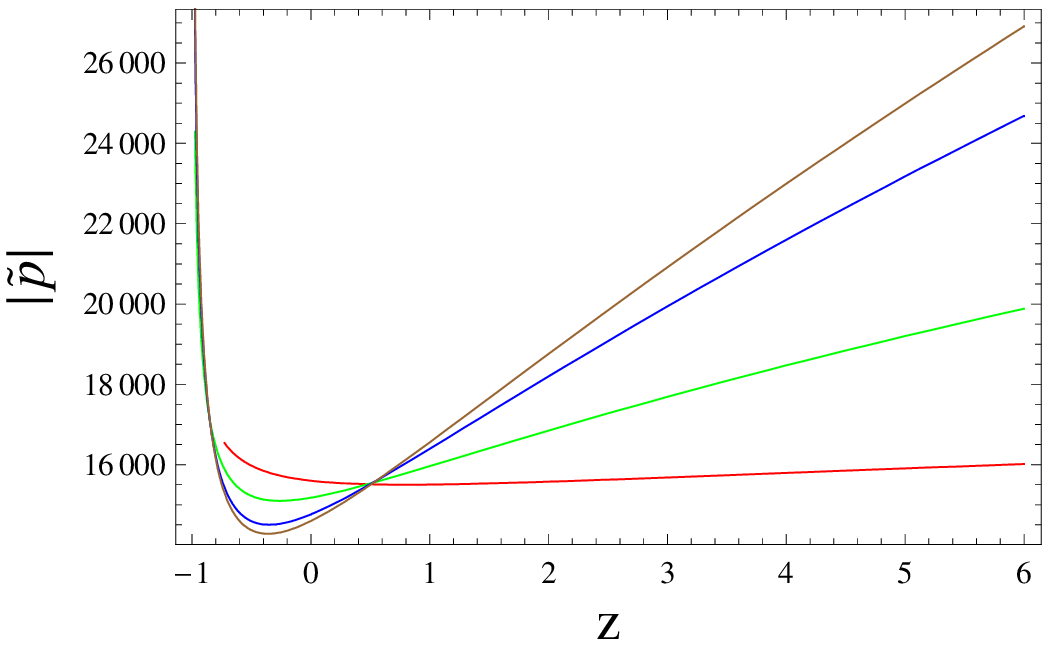}
  \caption{\label{figptilda}Plot of absolute value of effective pressure $|\tilde{p}|=|p_{DE}+p_{\nu}+\Pi_{DE}+\Pi_{\nu}|$ based\\ on Eq.(\ref{ptilda}).}
  \end{minipage}%
\begin{minipage}{.5\textwidth}
  \centering
  \includegraphics[width=18pc]{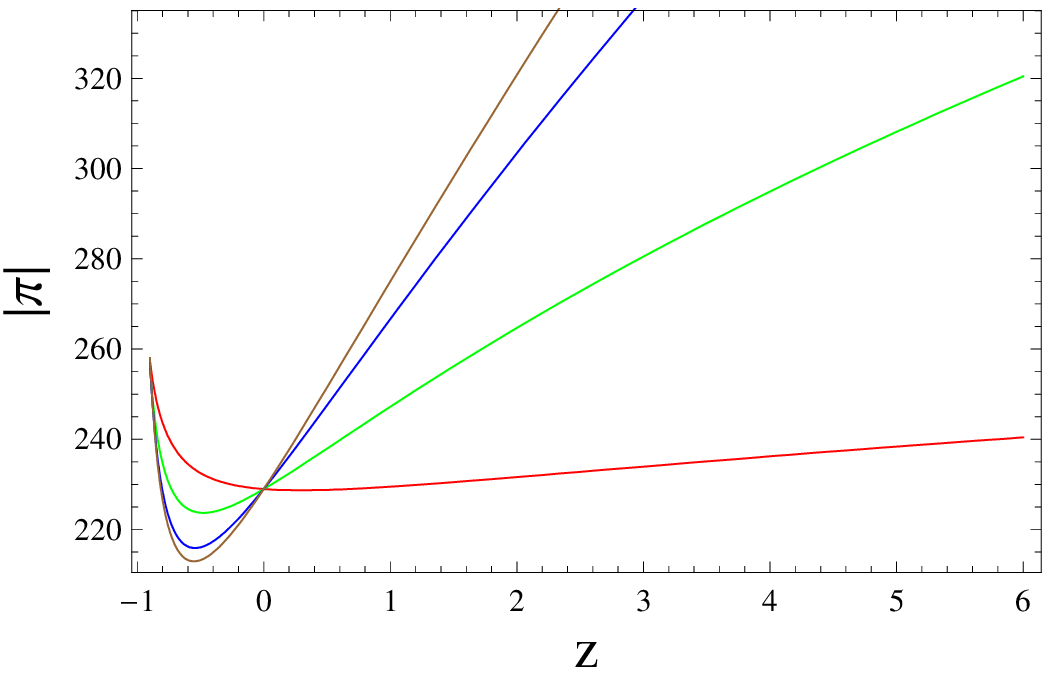}
  \caption{\label{figabspi}Plot of absolute value of the bulk viscous pressure $|\Pi|=|\Pi_{DE}+\Pi_{\nu}|$. }
 \end{minipage}
\end{figure}

\begin{figure}
\begin{minipage}{.5\textwidth}
  \centering
  \includegraphics[width=18pc]{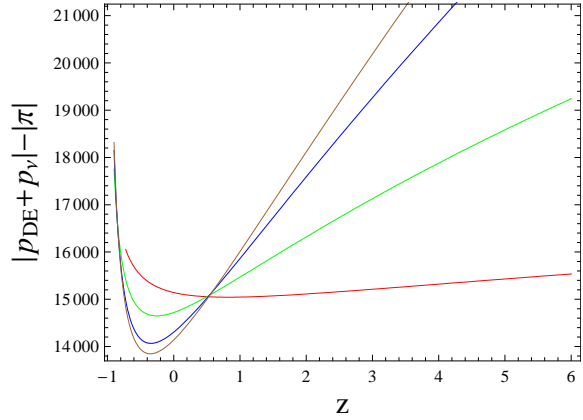}
  \caption{\label{figdiff}Plot of the difference between $|\Pi|=|\Pi_{DE}+\Pi_{\nu}|$ and $|p_{DE}+p_{\nu}|$. }
 \end{minipage}
\end{figure}

In Table \ref{tabeos} we have computed different values of $w_{total}$ and $w_{DE}$ for the current universe $(z=0)$ and for different choices of $\nu$. Studying observational data sets from SNLS3, BAO and Planck+WMAP9+WiggleZ measurements of matter power spectrum \cite{bits} have fixed observational constraint on the EoS parameter for the current universe at $-1.06_{-0.13}^{+0.11}$ and found the best fit value of $w_{DE0}$ to be $-1.01$. Comparing our results with that of \cite{bits} we observed that the EoS parameter due to viscous EHRDE is well within the range specified by \cite{bits} for all values of $\nu$. Moreover, $w_{DE0}(=-1.0018)$ is almost equal to the best fit value i.e. -1.01 for $\nu=0.15$. Thus, $\Lambda$CDM scenario is expected to be favoured by the viscous EHRDE considered here.

In Fig.\ref{figptilda} we have plotted $|\tilde{p}|=|p_{DE}+p_{\nu}+\Pi_{DE}+\Pi_{\nu}|$ based on Eq.(\ref{ptilda}). This figure shows that the magnitude of the effective pressure is decreasing with the expansion of the universe and it may be noted that this decrease is occurring till $z\approx -0.4$. In Fig.\ref{figabspi} we have plotted the magnitude of bulk viscous pressure $\Pi=\Pi_{DE}+\Pi_{\nu}$. Comparing Figs. \ref{figptilda} and \ref{figabspi} we can interpret that both $|\tilde{p}|$ and $|\Pi|$ have approximately similar decaying pattern. Although the rate of decrease in magnitude of $|\Pi|$ is higher than $|\tilde{p}|$, it may be noted that this pattern wise similarity indicates a significant contribution of the bulk viscosity to the effective pressure. Secondly, we further observe from Fig. \ref{figdiff} that $|p_{DE}+p_{\nu}|-|\Pi|\gg 0$ i.e. $|\Pi|\ll |p_{DE}+p_{\nu}|$. This implies that the non-equilibrium bulk viscous pressure is significantly smaller than the local equilibrium pressure \cite{vcg}.

To further consolidate observations made through EoS parameter a pair of cosmological parameters $\{r,s\}$, the so-called
``statefinder parameters", introduced by Sahni et al. \cite{state1} and and Alam et al. \cite{state2} used to discriminate between the various candidates of dark energy. If the $\{r-s\}$ trajectory meets the point $\{r=1,s=0\}$ then the model is said to attain $\Lambda$CDM phase of the universe. Statefinder parameters for different dark energy candidates have been studied in \cite{state3,state4,state5,state6}. The $\{r,s\}$ parameters are given by
\begin{eqnarray}
r &=& q+2q^2+\frac{\dot{q}}{H}\label{r1} \\
s &=& \frac{r-1}{3\left(q-\frac{1}{2}\right)}. \label{s1}
\end{eqnarray}
where, $q$ is the deceleration parameter. Hence, in the current framework Eq.(\ref{r1}) and (\ref{s1}) take the form
\begin{equation}
\begin{array}{c}\label{r}
r=-1+\frac{1}{4} \left(-6 (-1+\alpha ) \nu +6 \sqrt{3} \tau _{\text{DE}}-\Delta ^+ \left(2 \alpha +3 \beta  \nu +\beta  \Delta ^+\right)-\frac{C_1
e^{x \Delta ^-} \left(\Delta ^--\Delta ^+\right) \left(2 \alpha +3 \beta  \nu +\beta  \left(\Delta ^-+\Delta ^+\right)\right)}{C_1 e^{x \Delta
^-}+e^{x \Delta ^+} \left(-C_1+H_0^2\right)}\right.\\
\left.-\frac{C_1 e^{x \Delta ^-} \Delta ^- \left(\Delta ^--\Delta ^+\right) \left(2 \alpha +3 \beta
 \nu +\beta  \left(\Delta ^-+\Delta ^+\right)\right)}{C_1 e^{x \Delta ^-}+e^{x \Delta ^+} \left(-C_1+H_0^2\right)}+\right.\\
\left.\frac{C_1 e^{x \Delta ^-} \left(\Delta ^--\Delta ^+\right) \left(C_1 e^{x \Delta ^-} \Delta ^-+e^{x \Delta ^+} \left(-C_1+H_0^2\right)
\Delta ^+\right) \left(2 \alpha +3 \beta  \nu +\beta  \left(\Delta ^-+\Delta ^+\right)\right)}{\left(C_1 e^{x \Delta ^-}+e^{x \Delta ^+} \left(-C_1+H_0^2\right)\right){}^2}+\right.\\
\left.\frac{1}{2}
\left(4+6 (-1+\alpha ) \nu -6 \sqrt{3} \tau _{\text{DE}}+\Delta ^+ \left(2 \alpha +3 \beta  \nu +\beta  \Delta ^+\right)+\frac{C_1 e^{x \Delta
^-} \left(\Delta ^--\Delta ^+\right) \left(2 \alpha +3 \beta  \nu +\beta  \left(\Delta ^-+\Delta ^+\right)\right)}{C_1 e^{x \Delta ^-}+e^{x
\Delta ^+} \left(-C_1+H_0^2\right)}\right)^2\right)\\
\end{array}
\end{equation}

\begin{equation}
\begin{array}{c}\label{s}
s=8-\left(6 (1-\alpha ) \nu +6 \sqrt{3} \tau _{\text{DE}}-\Delta ^+ \left(2 \alpha +3 \beta  \nu +\beta  \Delta ^+\right)-\right.\\
\left.\frac{C_1 e^{x \Delta
^-} \left(\Delta ^--\Delta ^+\right) \left(2 \alpha +3 \beta  \nu +\beta  \left(\Delta ^-+\Delta ^+\right)\right)}{C_1 e^{x \Delta ^-}+e^{x
\Delta ^+} \left(-C_1+H_0^2\right)}-\frac{C_1 e^{x \Delta ^-} \Delta ^- \left(\Delta ^--\Delta ^+\right) \left(2 \alpha +3 \beta  \nu
+\beta  \left(\Delta ^-+\Delta ^+\right)\right)}{C_1 e^{x \Delta ^-}+e^{x \Delta ^+} \left(-C_1+H_0^2\right)}+\right.\\
\left.\left.\frac{C_1 e^{x \Delta ^-} \left(\Delta ^--\Delta ^+\right) \left(C_1 e^{x \Delta ^-} \Delta ^-+e^{x \Delta ^+} \left(-C_1+H_0^2\right)
\Delta ^+\right) \left(2 \alpha +3 \beta  \nu +\beta  \left(\Delta ^-+\Delta ^+\right)\right)}{\left(C_1 e^{x \Delta ^-}+e^{x \Delta ^+} \left(-C_1+H_0^2\right)\right){}^2}+\right.\right.\\
\left.\left.\frac{1}{2}
\left(4+6 (-1+\alpha ) \nu -6 \sqrt{3} \tau _{\text{DE}}+\Delta ^+ \left(2 \alpha +3 \beta  \nu +\beta  \Delta ^+\right)+\frac{C_1 e^{x \Delta
^-} \left(\Delta ^--\Delta ^+\right) \left(2 \alpha +3 \beta  \nu +\beta  \left(\Delta ^-+\Delta ^+\right)\right)}{C_1 e^{x \Delta ^-}+e^{x
\Delta ^+} \left(-C_1+H_0^2\right)}\right){}^2\right)\right/\\
\left(3 \left(6+6 (-1+\alpha ) \nu -6 \sqrt{3} \tau _{\text{DE}}+\Delta ^+ \left(2 \alpha +3 \beta  \nu +\beta  \Delta ^+\right)+\frac{C_1
e^{x \Delta ^-} \left(\Delta ^--\Delta ^+\right) \left(2 \alpha +3 \beta  \nu +\beta  \left(\Delta ^-+\Delta ^+\right)\right)}{C_1 e^{x \Delta
^-}+e^{x \Delta ^+} \left(-C_1+H_0^2\right)}\right)\right)
\end{array}
\end{equation}

\begin{figure}
\centering
\begin{minipage}{.5\textwidth}
  \centering
  \includegraphics[width=18pc]{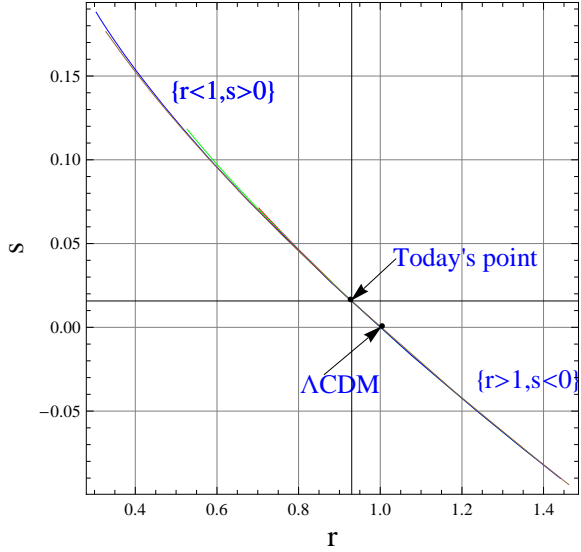}
  \caption{\label{rs}The time evolution of the $\{r,s\}$ trajectory.}
  \label{figeosD}
\end{minipage}%
\begin{minipage}{.5\textwidth}
  \centering
  \includegraphics[width=18pc]{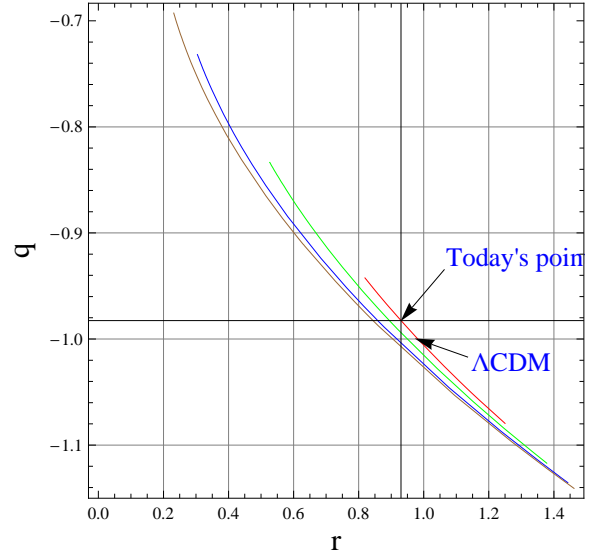}
  \caption{\label{qr}The time evolution of the $\{r,q\}$ trajectory. }
  \label{figeos}
\end{minipage}
\end{figure}

\begin{table}[h!]
  \centering
  \caption{Values of statefinder parameters at $z=0$ for different values of $\nu$.}
  \label{tabrs}
  \begin{tabular}{c|c|c|c|c}
    \hline\hline
    Statefinder parameters & $\nu=0.15$ & $\nu=0.20$ & $\nu=0.25$ & $\nu=0.27$\\
    \hline
   $r$ & 0.930117& 0.775948& 0.6248& 0.564796\\
    \hline
   $s$ &0.0157118& 0.0518002& 0.0892252& 0.104685\\
   \hline
   $q$ &-0.982605& -0.941773& -0.901696& -0.885755\\
   \hline
  \end{tabular}
\end{table}
In Fig.\ref{rs} we plot the time evolution of $\{r-s\}$ trajectory and observe that the $\Lambda$CDM fixed point i.e. $\{r=1,s=0\}$ is attainable by the statefinder trajectory. Moreover, it is worth noting that crossing the $\Lambda$CDM fixed point the trajectory is reaching the fourth quadrant i.e. $\{r>1,s<0\}$. Table \ref{tabrs} shows that location today's point is $\{r=0.881506,s=0.026874\}$ for $\nu=0.25$ and for $\nu=0.25$ this point is located very close to the $\{r=1,s=0\}\mid_{\Lambda CDM}$ fixed point. For other values of $\nu$, location of today's point is situated at larger distance from the $\{r=1,s=0\}$ point. Thus, it is observed that variation in the value of $\nu$ has a significant impact on the extent of separation of the current viscous EHRDE and increase in $\nu$ takes the model away from $\Lambda$CDM, where $\alpha$ and $\beta$ are fixed at $0.97$ and $0.37$ respectively.

We consider the case $\nu=1$, which corresponds to the EoS of dark matter $w_m=0$. In the situation of coexistence of viscous EHRDE and viscous dark matter the current value of statefinder pair is $\{r=-1.5428,~1.04588\}$ and it far away from the $\Lambda$CDM fixed point. However, in late time universe the trajectory reaches $\{r=1,~s=0\}$. Considering the smaller values $\nu=0.15,~0.20,~0.25$ and $0.27$ it is noteworthy that for different values of $\nu$ evolution of the $\{r,s\}$ trajectory starts from different points in the region $\{r<1,s>0\}$ and for $\nu=0.15$ the evolution begins from a point closer to $\Lambda$CDM fixed point compared to the higher values of $\nu$. However, all the trajectories after passing through their current values (see Table \ref{tabrs}) converge to the $\Lambda$CDM point in late time. It may further be noted that for $\nu=0.15$ the today's ($z=0$) point is very close to $\Lambda$CDM point. Therefore, although $\Lambda$CDM is favoured by the EoS parameter, the viscous EHRDE along with barotropic fluid can be discriminated (although very close) from $\Lambda$CDM point through statefinder parameters. It may be further noted that the evolution of the trajectories are not ending at $\{r=1,~s=0\}$, rather they are going to the region $r>1,~s<0$. Before reaching $\{r=1,~s=0\}$ the trajectories are traversing through the region $\{r<1,s>0\}$ only.

In Fig.\ref{qr} we consider the evolution of $\{r,q\}$ trajectory, where $q$ is the deceleration parameter. For all the values of $\nu$ the $\{r,q\}$ trajectory is starting its evolution from $\{r<1,q>-1\}$. For $\nu=0.15$ the trajectory is reaching $\{r=1,q=-1\}$ i.e. $\Lambda$CDM is attainable by the model. However, the today's ($z=0$) point is not coincident (although) very close to this fixed point. This further supports the observations through $\{r-s\}$ trajectory. Moreover, for $\Lambda$CDM the trajectory ends at $\{r=1,q=-1\}$, whereas for the current model the trajectory is going beyond that. The case $q<-1$ indicates a super accelerated scenario. For $\nu=0.15$, we find that $\{r=0.930117,q=-0.982605\}$ (see Table \ref{tabrs}) that is very close to $\Lambda$CDM.

\section{Perturbation equations}
We consider the linear perturbation of the current viscous EHRDE model towards a dark energy dominated universe. For this purpose, squared speed of sound
$c_s^2=\frac{d p_{total}}{d \rho_{total}}$ is crucial. A negative $c_s^2$ implies classical instability of a given perturbation. Myung \cite{myung} used squared speed of sound to discriminate between holographic dark energy, Chaplygin gas, and tachyon model and found a negative squared speed for holographic dark energy under the assumption of future event horizon as the IR-cutoff. In the present case squared speed of sound takes the form
\begin{equation}
\begin{array}{c}\label{csqr}
c_s^2=\frac{\dot{p}_{total}}{\dot{\rho}_{total}}=\\
-1+\frac{1}{6} \left(6 (1-\alpha ) \nu +6 \sqrt{3} \tau _{\text{DE}}-\Delta ^+
\left(2 \alpha +3 \beta  \nu +\beta  \Delta ^+\right)-\frac{C_1
e^{x \Delta ^-} \Delta ^- \left(\Delta ^--\Delta ^+\right) \left(2 \alpha +3 \beta  \nu +\beta  \left(\Delta ^-+\Delta ^+\right)\right)}{C_1
e^{x \Delta ^-} \Delta ^--e^{x \Delta ^+} \left(C_1-H_0^2\right) \Delta ^+}\right)
\end{array}
\end{equation}
\begin{figure}
\begin{minipage}{.5\textwidth}
  \centering
  \includegraphics[width=18pc]{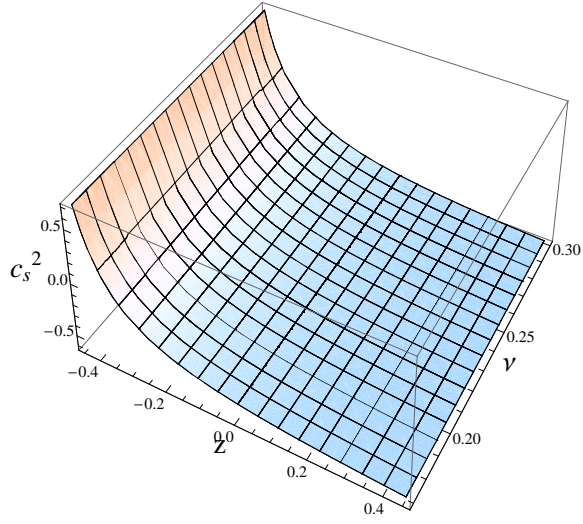}
  \caption{\label{figsound}Plot of $c_s^2$ based on Eq.(\ref{csqr}). }
 \end{minipage}
\end{figure}
In Fig.\ref{figsound} the squared speed of sound as derived in Eq.(\ref{csqr}) is plotted against redshift $z$ for a range of values of $\nu$. It is observed that $|c_s^2|<1$ and up to $z\approx -0.3$ we have $c_s^2<0$. This implies that from the early to some later stage the universe is unstable against small perturbation. In other words this negative $c_s^2$ indicates that even small perturbation can grow with time leading to an unstable universe. However, for $z\lesssim -0.3$ we are seeing $c_s^2>0$. Thus, at this stage the universe has gained stability. Hence, it may be interpreted that a viscous EHRDE along with viscous barotropic fluid can lead to a stable universe in very late stage.

\begin{figure}
\includegraphics[width=18pc]{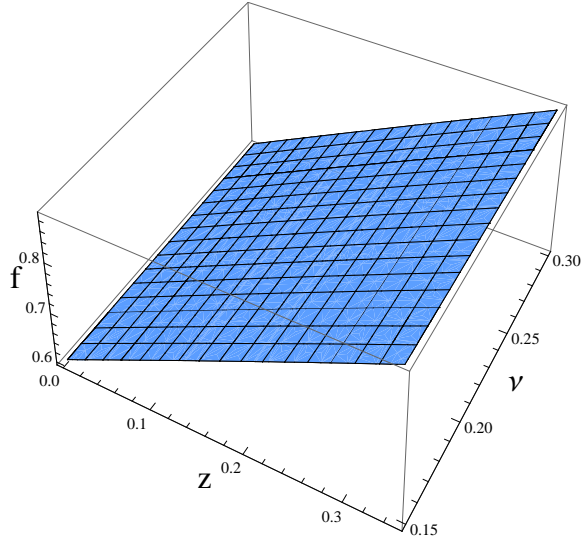}
\caption{\label{figgrowth} Evolution of growth factor $f=\frac{d \log \delta}{d \log a}$.  }
\end{figure}

Matter density perturbation given by $\delta=\frac{\delta \rho_m}{\rho_m}$, where fluctuation in matter density is given by $\delta \rho_m$. In the linear regime the perturbation $\delta$ satisfies \cite{arXiv}
\begin{equation}\label{delta}
\ddot{\delta}+2H\dot{\delta}-4\pi G\rho_m\delta=0
\end{equation}
Growth function is given by $f=\frac{d \log \delta}{d \log a}$.
\begin{figure}[h]
\begin{minipage}{.5\textwidth}
\includegraphics[width=18pc]{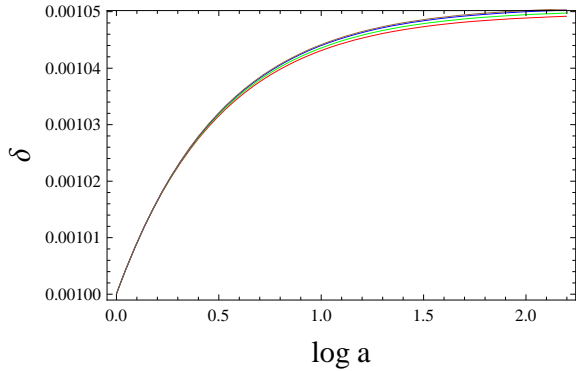}
\caption{\label{figdelta} Evolution of matter density perturbation\\ $\delta$ based on Eq.(\ref{diffdelta}) with scale factor.  }
\end{minipage}%
\end{figure}
Considering $\rho_m=\rho_{\nu}$ as available in Eq.(\ref{rhonu}) and $H$ as in Eq.(\ref{ysquare}) the Eq. (\ref{delta}) takes the form
\begin{equation}
\begin{array}{c}\label{diffdelta}
2 \left(C_1 e^{x \Delta ^-}+(H_0^2-C_1) e^{x \Delta ^+}\right)  \delta ''(x)+\\
 \left(C_1 e^{x \Delta ^-} \left(4+\Delta ^-\right)+(H_0^2-C_1) e^{x \Delta ^+} \left(4+\Delta ^+\right)\right) \delta'(x)
-\delta(x) =0
\end{array}
\end{equation}
Eq.(\ref{diffdelta}) is numerically solved and $\delta$ is plotted against $x=\log a$ in Fig.\ref{figdelta}. It is observed that $|\delta|<<1$, which indicates linear growth of fluctuations. In Fig.\ref{figgrowth} we have plotted the growth function $f=\frac{d\log \delta}{d\log a}$ for a range of values of $\nu$. This figure indicates that $f$ is a decreasing function of $z$. We observe that for $z=0.35$ the $f\thickapprox 0.88$, which is consistent with \cite{perturbation} and $z=0.22$ the $f\thickapprox 0.72$, which is consistent with \cite{perturbation1}.

\section{Second law of thermodynamics}

Discovery of black hole thermodynamics in 1973 \cite{bek} prompted physicists to study the thermodynamics of cosmological models of the universe \cite{thermobamba,setarethermo,th3,th4,th5,th6,th7}. Bekenstein \cite{bek}associated event horizon
and the thermodynamics of a black hole by showing that event horizon of the black
hole is a measure of the entropy of it. In subsequent studies this idea has been generalized to horizons of
cosmological models by connecting each horizon to an entropy \cite{setarethermo}. This modified the second
law of thermodynamics to its generalized form, in which the time derivative of the total entropy i.e. the sum of the time derivative of the entropy on the horizon and the fluid inside the horizon must be non-negative i.e. $\dot{S}_{total}\geq 0$. In the present work we consider event horizon $R_E$ as the enveloping horizon of the universe. The event horizon is given by \cite{setarethermo}
\begin{equation}\label{event}
R_E=a\int_t^{t_s}\frac{dt}{a}=a\int_a^{t_s}\frac{da}{Ha^2}
\end{equation}
where for different spacetimes $t_s$ has different values, e.g. for de
Sitter space–time $t_s=\infty$. The event horizon satisfies
\begin{equation}\label{diff}
\dot{R}_E=HR_E-1
\end{equation}
Taking reconstructed $H$  (Eq.(\ref{ysquare})) in Eq.(\ref{diff}) we get the solution for $R_E$ as
\begin{equation}
R_E=e^x C_3+\frac{2 \sqrt{1+\frac{C_1 e^{x \left(\Delta ^--\Delta ^+\right)}}{(H_0^2-C_1)}} 2F1\left[-\frac{2+\Delta
^+}{2 \Delta ^--2 \Delta ^+},\frac{1}{2},1-\frac{2+\Delta ^+}{2 \Delta ^--2 \Delta ^+},-\frac{C_1 e^{x \left(\Delta ^--\Delta ^+\right)}}{(H_0^2-C_1)}\right]}{\left(2+\Delta
^+\right)\sqrt{C_1 e^{x \Delta ^-}+(H_0^2-C_1) e^{x \Delta ^+}} }
\end{equation}
Hawking temperature on the horizon is \cite{sir}
\begin{equation}
T_E=\frac{H^2R_E}{2\pi}
\end{equation}
and subsequently entropy on the event horizon is \cite{sir}
\begin{equation}\label{on}
\dot{S}_E=\frac{8\pi^2R_E^2(p_{total}+\rho_{total})}{H}
\end{equation}
To determine the entropy variation of the fluid inside $R_h$ we start with the Gibb’s
relation
\begin{equation}\label{inside}
T_f dS_f=dE+p_{total}dV
\end{equation}
where, volume of the fluid is $V=\frac{4}{3}\pi R_E^3$ and total energy of the fluid is $E=\rho_{total}V$.
\begin{figure}[h]
\includegraphics[width=20pc]{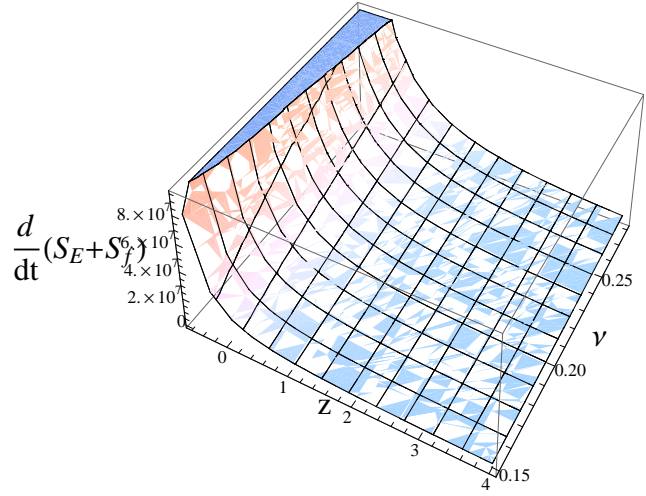}
\caption{\label{figthermo} Time derivative of total entropy $\dot{S}_{total}$.}
\end{figure}
Using Eqs.(\ref{rhode}), (\ref{rhonu}) and (\ref{pde}) in Eqs. (\ref{on}) and (\ref{inside}) we get the time variation of total entropy as
\begin{equation}
\begin{array}{c}
\dot{S}_E+\dot{S}_{f}=\frac{8 \pi ^2}{\left(C_1 \left(e^{x \Delta ^-}-e^{x \Delta ^+}\right)+e^{x \Delta ^+} H_0^2\right){}^{3/2} \left(2+\Delta ^+\right)^2}
 \left(C_1 e^{x \Delta ^-} \Delta ^--e^{x \Delta ^+} \left(C_1-H_0^2\right) \Delta ^+\right) \\
\left(\left(2+\Delta ^+\right) \left(2\times 2F1\left[\frac{1}{2},-\frac{2+\Delta ^+}{2 \Delta ^--2 \Delta ^+},1-\frac{2+\Delta
^+}{2 \Delta ^--2 \Delta ^+},\frac{C_1 e^{x \left(\Delta ^--\Delta ^+\right)}}{C_1-H_0^2}\right] \sqrt{-\frac{C_1 \left(-1+e^{x
\left(\Delta ^--\Delta ^+\right)}\right)+H_0^2}{C_1-H_0^2}}+\right.\right.\\
\left.\left.2 C_3 e^x \sqrt{C_1 \left(e^{x \Delta ^-}-e^{x \Delta ^+}\right)+e^{x
\Delta ^+} H_0^2}+C_3 e^x \sqrt{C_1 \left(e^{x \Delta ^-}-e^{x \Delta ^+}\right)+e^{x \Delta ^+} H_0^2} \Delta ^+\right)-\right.\\
\left.\left(2\times 2F1\left[-\frac{2+\Delta ^+}{2 \Delta ^--2 \Delta ^+},\frac{1}{2},1-\frac{2+\Delta ^+}{2 \Delta ^--2 \Delta
^+},\frac{C_1 e^{x \left(\Delta ^--\Delta ^+\right)}}{C_1-H_0^2}\right] \sqrt{-\frac{C_1 \left(-1+e^{x \left(\Delta ^--\Delta ^+\right)}\right)+H_0^2}{C_1-H_0^2}}+\right.\right.\\
\left.\left.2C_3 e^x \sqrt{C_1 \left(e^{x \Delta ^-}-e^{x \Delta ^+}\right)+e^{x \Delta ^+} H_0^2}+C_3 e^x \sqrt{C_1 \left(e^{x \Delta
^-}-e^{x \Delta ^+}\right)+e^{x \Delta ^+} H_0^2} \Delta ^+\right){}^2\right)
\end{array}\label{sdottotal}
\end{equation}
\section{Concluding Remarks}
In the present work we have extended the study of \cite{feng} on viscous holohraphic Ricci dark energy to extended holographic Ricci dark energy (EHRDE) under the influence of bulk viscosity. Considering a coexistence of viscous EHRDE and viscous barotropic fluid with equation of state $p_{\nu}=(\nu-1)\rho_{\nu}$ we have reconstructed Hubble parameter $H$ (see Eq. (\ref{ysquare})) corresponding to $\nu=0.15,~0.20,~0.25$ and $0.27$ respectively. The reconstructed Hubble parameter is found to be decreasing with evolution of the universe and in a later stage $-0.7\lesssim z\lesssim -0.25$, the reconstructed $H$ is found to start increasing. Thus, in a later stage $\dot{a}(t)$ has been found to have the possibility of dominating $a(t)$. Equation of state (EoS) parameter for the viscous EHRDE based on Eq.(\ref{eosD}) has been observed to exhibit a transition from $EoS>-1$ (quintessence) to $EoS<-1$ (phantom) at $z\approx 0.01$ i.e. in an earlier stage of the universe and the current value of the EoS parameter for viscous EHRDE is found to favour the $\Lambda$CDM model and the $w_{DE0}$ for the current model is consistent with results obtained by \cite{bits} through observational data sets from SNLS3, BAO and Planck+WMAP9+WiggleZ measurements. It has further been noted that irrespective of the values of $\nu$ the viscous EHRDE is behaving like ``quintom" i.e. transiting from quintessece to phantom. It has also been observed that the behaviour of $w_{total}$ (see Fig.\ref{figeostotal} ) is largely similar to that of $w_{DE}$ as far as the ``quintom" behaviour is concerned. However, the transition to phantom has been found to occur in a later stage $z\approx -0.6$ for all values of $\nu$. To further consolidate the results from EoS study we have studied the statefinder diagnostics of the viscous dark energy-barotropic fluid model by deriving expressions for the $\{r-s\}$ parameters (see Eqs. (\ref{r}) and (\ref{s})) for the current model. Plotting $\{r-s\}$ for different values of $\nu$ the trajectories are found to pass through today's point ($z=0$) (see Table \ref{tabrs}) and subsequently converge to the $\Lambda$CDM point in a later stage. It has been noted that for $\nu=0.15$ the today's point is very close to $\Lambda$CDM point. From this observation it may be interpreted that although $\Lambda$CDM is favoured by the EoS parameter, the current model can be discriminated (although very close) from $\Lambda$CDM point through statefinder parameters. Moreover, evolution of the trajectories have not ended at $\{r=1,~s=0\}$, rather they reached the region $r>1,~s<0$. Similar behaviour has been observed through a study of $\{r-q\}$ trajectory too along with attainment of a super-accelerated phase of the universe in a later stage. In Fig. \ref{figdiff} we have observed $|p_{DE}+p_{\nu}|-|\Pi|\gg 0$ i.e. $|\Pi|\ll |p_{DE}+p_{\nu}|$, which implies that the non-equilibrium bulk viscous pressure is significantly smaller than the local equilibrium pressure.

In the next phase we have studied behaviour of squared speed of sound $c_s^2$ to study the stability of the model. After obtaining the expression for $c_s^2$ for the current model in Eq. (\ref{csqr}) and plotting in Fig.\ref{figsound} we have observed that currently $c_s^2<0$ indicating instability of the model under small perturbations. However, for $z\lesssim -0.3$ it has been observed that $c_s^2>0$ indicating that at this stage the universe has gained stability. It has been interpreted from this result that the current model can lead to a stable universe in very late stage. We have also studied the matter density perturbation given by $\delta=\frac{\delta \rho_m}{\rho_m}$ and the growth function $f$ for the current model. Referring to Fig.\ref{figdelta} it is observed that $|\delta|<<1$ indicating linear growth of fluctuations. The growth function $f=\frac{d\log \delta}{d\log a}$ plotted in Fig.\ref{figgrowth} for a range of values of $\nu$ it has been observed that $f$ is a decreasing function of $z$. It has been noted that for $z=0.35$ the $f\thickapprox 0.88$, which is consistent with \cite{perturbation} and for $z=0.22$ the $f\thickapprox 0.72$, which is consistent with \cite{perturbation1}.

Finally we have considered the generalized second law of thermodynamics (GSL) for the current model considering event horizon as the enveloping horizon of the universe. Under the assumption thermal equilibrium we have observed that time derivative of total entropy $\dot{S}_{total}\geq 0$ throughout the evolution of the universe (see Fig.\ref{figthermo}) and hence the GSL is found valid in a universe containing EHRDE and barotropic fluid under the influence of viscosity.

\section{Acknowledgement}
Financial support from DST, Govt of India under project grant no. SR/FTP/PS-167/2011 is thankfully acknowledged.

\end{document}